\documentclass[runningheads]{llncs}
\usepackage{subfigure}     
\usepackage{graphicx}
\usepackage{multirow}
\usepackage{tabularx}
\usepackage{tabu}
\usepackage{amsmath}
\usepackage{amsfonts}
\usepackage{amssymb}
\usepackage{booktabs}       
\usepackage{amssymb}
\usepackage{color}
\usepackage{caption}
\usepackage[colorlinks,linkcolor=blue]{hyperref}

\begin{document}
\title{Towards Better Dermoscopic Image Feature Representation Learning for Melanoma Classification}

%
\author{ChengHui Yu \inst{1}
,
MingKang Tang \inst{1},
ShengGe Yang\inst{1},
MingQing Wang\inst{1},
Zhe Xu\inst{1}
JiangPeng Yan\inst{1,2},
HanMo Chen\inst{1},
Yu Yang\inst{1},
Xiao-Jun Zeng\inst{3},
Xiu Li\inst{1}
\thanks{C. Yu and M. Tang contributed equally. X. Li and J. Yan are corresponding authors.}}

\authorrunning{C. Yu, M. Tang, et al.}
\titlerunning{Better Image Feature Representation for Melanoma Classification}
%
\institute{
Tsinghua Shenzhen International Graduate School, Tsinghua University, Shenzhen 518055, Peoples R China \and Department of Automation, Tsinghua University, China \and 
Department of Computer Science, University of Manchester, U.K.\\\email{\{ych20,tmk20,ysg20,wmq20,xu-z18,yanjp17,chm20,yy20\}@mails.tsinghua.edu.cn,
x.zeng@manchester.ac.uk, li.xiu@sz.tsinghua.edu.cn}
}

\maketitle              

\begin{abstract}
Deep learning-based melanoma classification with dermoscopic images has recently shown great potential in automatic early-stage melanoma diagnosis. However, limited by the significant data imbalance and obvious extraneous artifacts, i.e., the hair and ruler markings, discriminative feature extraction from dermoscopic images is very challenging. In this study, we seek to resolve these problems respectively towards better representation learning for lesion features. Specifically, a GAN-based data augmentation (GDA) strategy is adapted to generate synthetic melanoma-positive images, in conjunction with the proposed implicit hair denoising (IHD) strategy. Wherein the hair-related representations are implicitly disentangled via an auxiliary classifier network and reversely sent to the melanoma-feature extraction backbone for better melanoma-specific representation learning. Furthermore, to train the IHD module, the hair noises are additionally labeled on the ISIC2020 dataset, making it the first large-scale dermoscopic dataset with annotation of hair-like artifacts. Extensive experiments demonstrate the superiority of the proposed framework as well as the effectiveness of each component. The improved dataset publicly avaliable at \href{https://github.com/kirtsy/DermoscopicDataset}{https://github.com/kirtsy/DermoscopicDataset}.

\keywords{Deep learning  \and Dermoscopic images \and Melanoma diagnosis \and Image classification.}
\end{abstract}

\section{Introduction}
Melanoma is a rare but highly fatal skin cancer, leading to nearly 75\% skin cancer associated deaths~\cite{melanoma_harmful}. Early detection and diagnosis of melanoma are crucial since the five-year survival rate will rise to 97\% if identified at a curable stage~\cite{survival_rate}. However, current methods for early melanoma diagnosis highly rely on clinical judgment on dermoscopic images. Due to the rareness of melanoma and its vagueness of early symptoms, the diagnosis results are highly dependent on physicians' clinical experience and diagnostic insights~\cite{melanoma_experience}. Consequently, nearly half of patients are diagnosed with late-stage cancer, and therefore suffer from pain by systemic anti-cancer therapy and accompanying side effects, which are entirely preventable~\cite{harmful_effect}. 

Recently, computer-aided diagnosis (CAD) has shown promising performance in the early diagnosis of cancer. With the CAD system, dermatologists are able to perform large-scale early detection of melanoma with human effort reduced. In particular, deep learning-based CAD becomes dominant on diagnostic accuracy by reason of its outstanding ability in extracting features from the input data~\cite{yan2021,xu2021,xu2021unsupervised,xu2021f3rnet}. To better exploit the input dermoscopic images, current melanoma classification models often rely on an ensemble strategy~\cite{ensemble_ahmed,ensemble1_Reisinho}. Despite their excellent performance in accuracy, such an ensemble strategy has various disadvantages, such as time consumption, resource demand, and interpretation difficulty. 

\begin{figure}
\vspace{-0.4cm}
\centering
\includegraphics[width=0.7\textwidth]{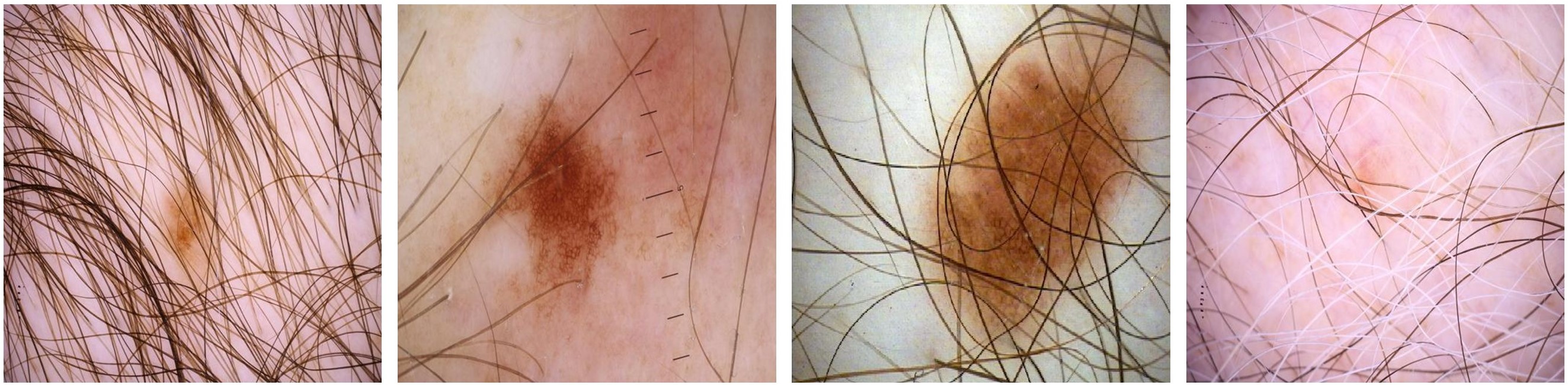}
\caption{Example cases with hair-like artifacts in the ISIC2020 dataset} \label{fig_noise}
\vspace{-0.3cm}
\end{figure}

In addition to the ensemble strategy, another rewarding approach towards satisfying outcomes is to make endeavors on more discriminative feature representation extraction from dermoscopic images. However, two main challenges are observed. Firstly, existing dermoscopic datasets are usually notably class-imbalanced because positive samples for melanoma are fewer than negative ones. For example, positive samples only account for 1.7\% of the entire ISIC2020 dataset~\cite{ISIC2020_data}. This substantial imbalance greatly hinders the representation learning of positive features. On the other hand, dermoscopic images are often accompanied by linear-shaped obstructions, i.e., hairs, as shown in Fig.\ref{fig_noise}. To eliminate such hair-like artifacts, some efforts are made to superficially erase them from images via segmentation mask-/morphological-based inpainting~\cite{GAN,hair_removal3}. However, this strategy may introduce other artifacts~\cite{xu2020adversarial} or undermine the relevant lesion pixels, thus impeding the subsequent feature extraction.

In this study, we focus on resolving the two aforementioned challenges, i.e., data imbalance and the hair-like artifact, towards better representation learning for lesion feature. Firstly, aiming to ameliorate the data imbalance problem, we adapt a GAN-based data augmentation (GDA) strategy to generate synthetic melanoma-positive images. As for hair denoising, our initial idea stems from the decision-making process of trained dermatologists. In a dermatologist-only situation, the clinicians can identify the subtle differences in local features regardless of the presence of hair and color differences based on their experience rather than using conventional bottom-up instant image analysis fashion~\cite{decision_making}. Correspondingly, we propose an implicit hair denoising (IHD) strategy, wherein the hair-related representations are implicitly disentangled via an auxiliary classifier network and reversely sent to the melanoma-feature extraction backbone for better melanoma-specific representation learning. Besides, in order to train the IHD module, the hair-like artifacts are additionally labeled on the ISIC2020 dataset~\cite{ISIC2020_data}, making it the first large-scale dermoscopy dataset with annotation of hair-like artifacts. Compared to the superficially inpainting-based methods \cite{GAN,hair_removal}, our approach focuses on the high-level feature representation without introducing other artifacts \cite{xu2020adversarial} to the images. Extensive experiments on multiple datasets demonstrate that our GDA and IHD strategies are effective with competitive results. The main contributions of this work are summarized as follows:

\begin{itemize}
\item[(1)] To address the notable data imbalance in the melanoma classification task, we adapt a GAN-based data augmentation strategy to generate synthetic melanoma-positive images.
\item[(2)] To eliminate the impact of hair-like artifacts, we present an implicit hair denoising strategy, where the hair-related representations are implicitly disentangled, facilitating more discriminative melanoma-specific representation learning.
\item[(3)] To train the implicit hair denoising module, we provide a melanoma dataset including 33,126 cases with manually annotated hair-like artifact labels upon the ISIC2020 dataset~\cite{ISIC2020_data}, making it the first large-scale dermoscopy dataset with annotation of hair-like artifacts.  
\end{itemize}

\section{Methods}

\begin{figure}[t]
\includegraphics[width=\textwidth]{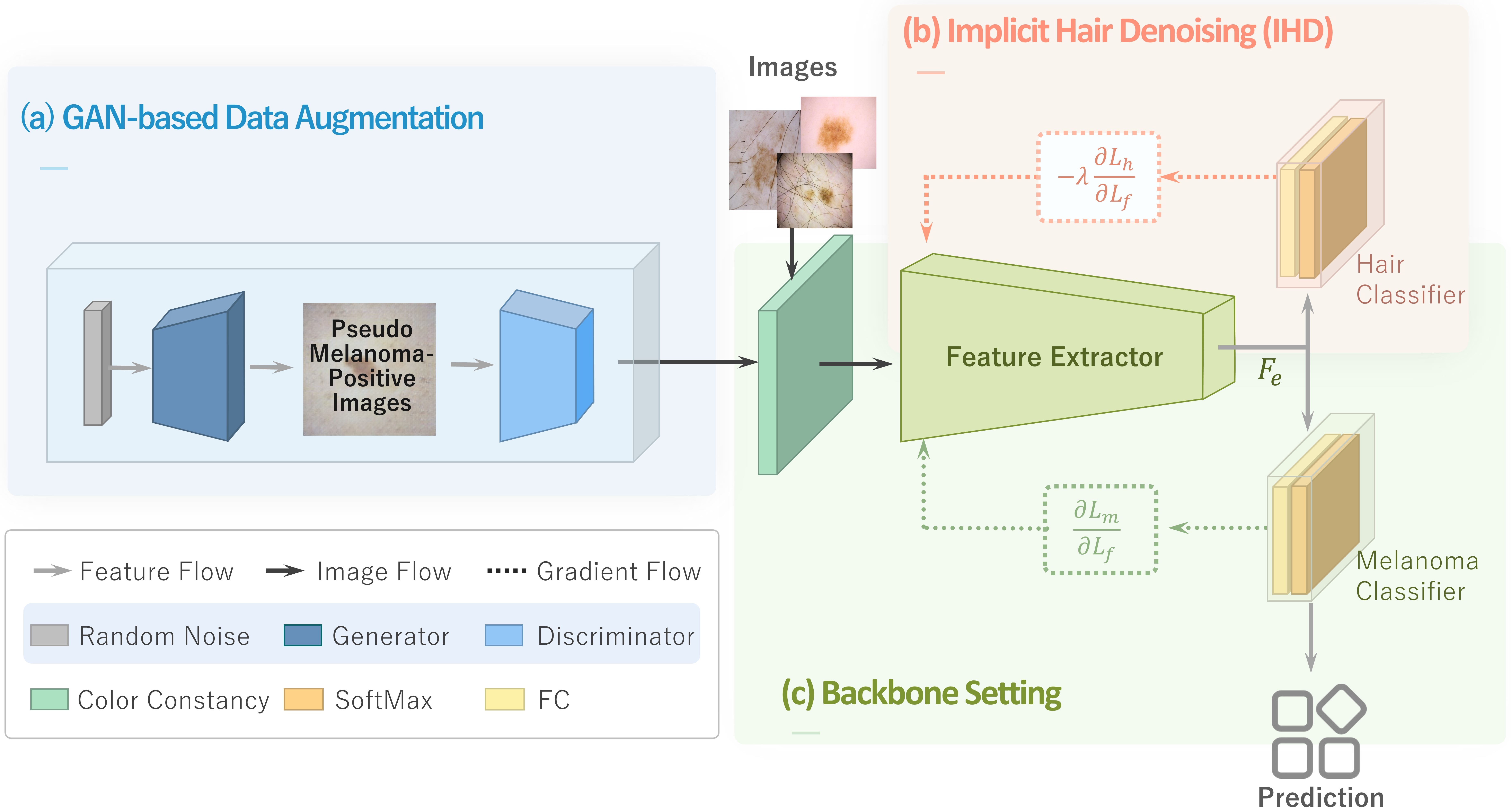}
\caption{The overall framework for melanoma diagnosis.} \label{fig1}
\vspace{-0.3cm}
\end{figure}

Fig.\ref{fig1} illustrates the framework of our proposed model for melanoma classification. The backbone consists of a color-constancy layer and a feature extractor to extract semantic features from dermoscopic images.Then our model classifies melanoma ONLY based on the features learned from dermoscopic images. We balance the class distribution of melanoma-positive/negative samples with the GDA strategy. Meanwhile, the hair-like artifacts would be implicitly eliminated via the IHD strategy. Details are elaborated as follows.


\subsection{Backbone Setting}
Our approach shares the standard backbone setting ~\cite{cc_main_network,Eff_b6_SOTA,DenNet_2_SOTA} consisting of a color constancy layer for image preprocessing and an encoder for feature extraction.

\subsubsection{Color constancy.} Color constancy is a typical step in dermoscopic image preprocessing to reduce the environmental impact. This step could possibly simulate the neuro-physiological mechanism of color perception. Comparing several existing algorithms (e.g., Shades of Gray~\cite{Sog} and Max-RGB~\cite{max_rgb}), we found that Max-RGB achieved superior performance on the ISIC2020 dataset~\cite{ISIC2020_data}. Thus, Max-RGB is adopted to adjust image colors in our study. Max-RGB assumes that white patches cause the maximum response of each color channel, which can be formulated as $\max \limits_{x} f(X)=ke$, where $f(X)=[R(X),B(X),G(X)]^\mathrm{T}$ represents the pixel values of the original image, $X$ is pixel coordinates, $k$ is calibration constant, and $e$ is illumination factor.

\subsubsection{Feature extraction.} Feature extraction via supervised learning is widely employed in dermoscopic image classification tasks~\cite{Eff_b6_SOTA,DenNet_2_SOTA}. Following the previous work in~\cite{meta_concat1}, we utilizes EfficientNet-B3~\cite{EffNet_b3} to extract the lesion features $F_e$, formulated by:

\begin{equation}
    \label{eq:1}
    F_e(x_{img};\mathbf{W}_e) = EfficientNet(x_{img}),
\end{equation}
where $\mathbf{W}_e$ represents a parametric matrix and $x_{img}$ denotes the input images after data enhancement (i.e., GDA and color constancy). $EfficientNet$ function indicates EfficientNet-B3. The backbone function $Backbone_m$ is parameterized by weights-biased softmax ($\mathbf{W}_m,b_m$): 
\begin{equation}
    \label{eq:6}
    Backbone_m(F_e;\mathbf{W}_m,b_m) = softmax(\mathbf{W}_mF_e+b_m).
\end{equation}

Then, the cross-entropy loss is used as the loss function in learning process:
 \begin{equation}
    \label{eq:7}
    \mathcal{L}_m(Backbone_m,y)=-\sum\log(Backbone_m)y,
\end{equation}
where $y$ represents the correct class label of input image $x_{img}$.

Formula (\ref{eq:8}) represents the objective function for the melanoma classifier optimization, where $i$ indicates the $i$-th sample.
\begin{equation}
    \label{eq:8}
    \min_{\mathbf{W}_e,\mathbf{W}_m,b_m} \left[ \dfrac{1}{n}\sum_{i=1}^N \mathcal{L}_m^i(Backbone_m,y) \right],
\end{equation}

Compared to regular CNN architecture, EfficientNet is deeper and wider with higher image resolution, allowing it to learn a broad range of clinically useful knowledge. EfficientNet is also generated by applying the compound scaling method, which leverages multi-objective neural architecture search that optimizes both accuracy and distribution of limited computation resources. Besides, we have tested several depths of EfficientNet, including B3, B4, and B6, and found that B3 is the optimized depth against the model's performance on ISIC2020~\cite{ISIC2020_data}.  


\subsection{GAN-based Data Augmentation Strategy}
\begin{figure}
\vspace{-0.5cm}
\includegraphics[width=\textwidth]{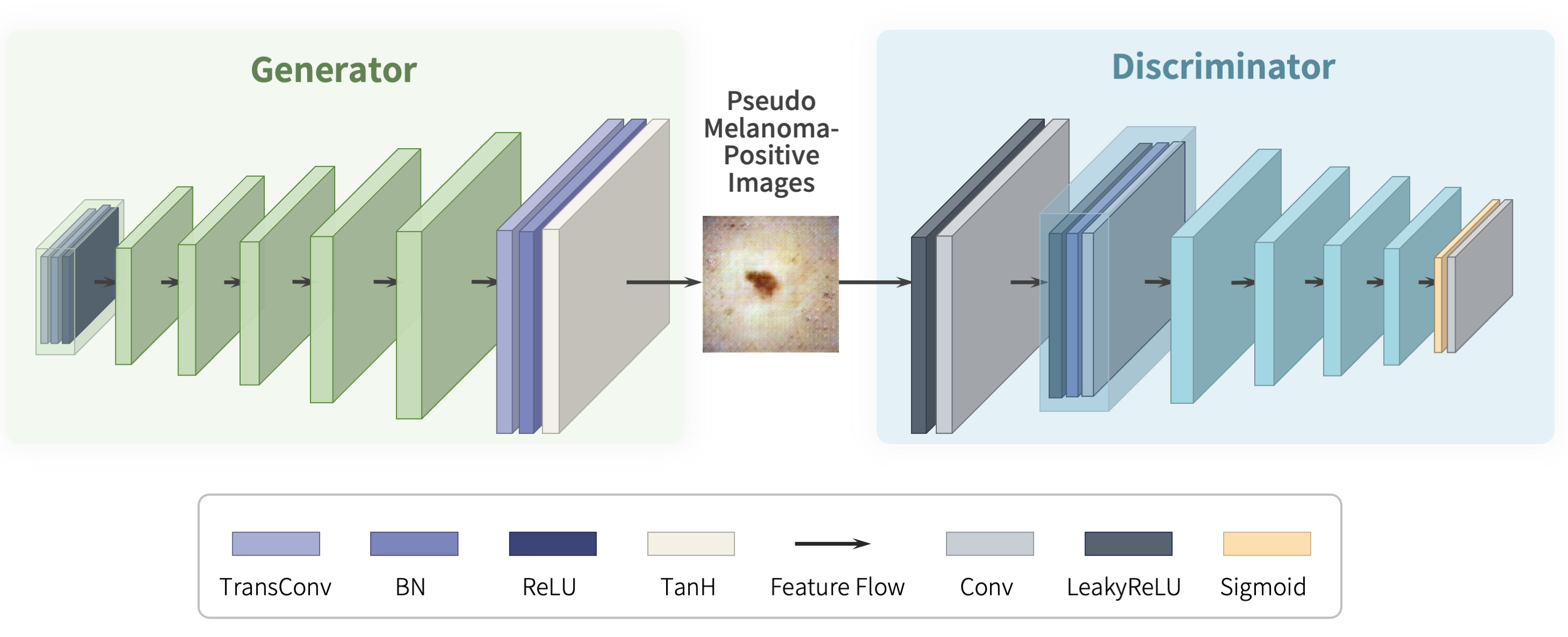}
\caption{The architecture of the decoupled DCGANs (Deep Convolutional Generative Adversarial Networks) used in our GAN-based data augmentation.} \label{fig2}
\vspace{-0.4cm}
\end{figure}

Due to the fact that melanoma is rare skin cancer, the positive and negative samples are severely imbalanced in the ISIC2020 dataset \cite{ISIC2020_data}, where there are 584 positive samples out of 33,126 images (only 1.76\%). Data augmentation is a promising strategy to cope with the severe data imbalance and improve the model generalizability. Inspired by~\cite{GAN},  decoupled DCGANs (as shown in Fig.\ref{fig2}) is adapted to generate extra 2,000 synthetic melanoma-positive images with the upsampled resolution of $512\times512$. DCGAN could be formulated as $\min \mathbf{G} \max{}_\mathbf{D} V(\mathbf{D},\mathbf{G})$, where $\mathbf{G}$ and $\mathbf{D}$ represent Generator and Discriminator respectively. $V(\mathbf{D},\mathbf{G})$ shows the difference between the synthetic and genuine samples. The generator minimizes the difference while the discriminator separates synthetic and genuine samples as much as possible. Specifically, the the generator is trained under the supervision of the discriminator to produce images with similar distribution as the ground-truth melanoma-positive images. The distribution difference is measured by the simultaneously trained Discriminator. Then, in order to maximize data variety, the generator compares each generated image with its closest training images, in which the minimum MSE is calculated to ensure non-repeatability. Besides, the wgan loss~\cite{wganloss} and RMSprop optimizer with a learning rate of 0.0002 are used to train both the generator and discriminator.


\subsection{Implicit Hair Denoising Strategy}
Hairs and other linear-shaped obstructions are usually present in dermoscopic images which result in noise in feature representation learning. Based on the fact that hair-like artifacts are often subconsciously neglected by dermatologists during diagnosis, the implicit hair denoising (IHD) strategy is proposed. Specifically, inspired by \cite{DANN} which demonstrates that joint optimization will lead to both discriminativeness and domain-invariance and thus augment a new gradient reversal layer while learning features, we integrate an auxiliary classifier network to implicitly disentangle the hair-related representations, as shown in Fig.\ref{fig1}(b). Then, the hair-related features are feedbacked to the melanoma-feature extraction backbone for better melanoma-specific representation learning. 

Formally, for EfficientNet-based features $F_e$ after feature extractor with a parametric matrix $\mathbf{W}_e$, the reversal layer function ${IHD}$ is parameterized by weights-biased softmax $(\mathbf{U}_h,v_h)$: 
\begin{equation}
\label{eq:2}
    IHD(F_e;\mathbf{U}_h,v_h) = softmax(\mathbf{U}_hF_e+v_h),
\end{equation}

Using the softmax function, each component of vector ${IHD}(F_e)$ represents that the neural network assigns $x_{img}$ to the conditional probability of the corresponding class. The cross-entropy between the correct hair label $y_h$ and ${IHD}$ is adopted as the loss function for classification:
\begin{equation}
    \label{eq:3}
    \mathcal{L}_h(IHD,y_h) = -\sum \log(IHD)y_h,
\end{equation}

Formula (\ref{eq:4}) is the optimization problem on the hair classifier training under the IHD strategy, where $i$ indicates the $i$-th sample.
\begin{equation}
    \label{eq:4}
    \max_{\mathbf{W}_e,\mathbf{U}_h,v_h}\left[ \dfrac{1}{n} \sum_{i=1}^N\mathcal{L}_h^i(IHD,y_h)  \right],
\end{equation}


\section{Experiments and Results} \label{Set_exp}
We evaluated our approach on two public dermoscopic datasets, ISIC2020~\cite{ISIC2020_data} and PH2~\cite{ph2}. The former was utilized to examine the feature extraction performance of our approach, while the latter was for generalization assessment of the model trained on the ISIC2020 dataset.

\subsection{Datasets and Implementation Details} \label{Sec_dataset}
The experimental dataset is generated from ISIC2020 dataset~\cite{ISIC2020_data}, the largest among recent releases, originally containing 33,126 endoscopic images with patients' metadata and lesion labels. In addition, we generated another 2,000 synthetic melanoma-positive images through the GDA to balance skew classes of samples. Furthermore, we manually labeled hair-like artifacts\footnote{The datasets generated and/or analyzed in this paper can be accessed from the corresponding author upon reasonable requests.} for the use of the IHD strategy. Five-folds cross-validation are adopted for model evaluation and selection after data preparation. The trained models are then fed into the dataset generated from PH2~\cite{ph2} for generalization evaluation.

The proposed model was implemented based on PyTorch and trained on four NVIDIA GeForce RTX 3090 GPUs using Adam optimizer with a learning rate of $\textit{3e-5}$. As suggested by equation (\ref{eq:8}) and equation (\ref{eq:4}), with the optimizer, the process of gradient descent can be formulated as follow:

\begin{equation}
    \label{eq:9}
    \begin{cases}
    \theta_m\leftarrow \theta_m - \eta\dfrac{g_{m,1}}{\sqrt{g_{m,2}}+\epsilon} \\
    \theta_h \leftarrow \theta_h - \eta\lambda\dfrac{g_{h,1}}{\sqrt{g_{h,2}}+\epsilon} \\
    \theta_f\leftarrow \theta_f - \eta\left( \dfrac{g_{m,1}}{\sqrt{g_{m,2}}+\epsilon} -\lambda \dfrac{g_{h,1}}{\sqrt{g_{h,2}}+\epsilon} \right) 
    \end{cases}
\end{equation}
where $\eta$ is the learning rate, and $\lambda$ is the weighted loss rate of hair-like artifacts. $g_{k,i} = \dfrac{\beta_i\theta_k+(1-\beta_i)\frac{\partial^i L_k}{\partial \theta_k^i}}{1-\beta_i}$ with subscript $k=\{m,h\}$ and $i=\{1,2\}$. Also, $g_{m,i}$ and $g_{h,i}$ correspondingly represent the melanoma and hair gradient.Where $\beta_1$ and $\beta_2$ denote the exponential decay rate of first-order and second-order moments, respectively. 

Moreover, we trained the IHD module without synthetic melanoma-positive images due to lack of hair-like artifact labelling. The batch size is set to 64, while input images are resized to $512\times512$ with 5-fold cross-validation. Color constancy is used in both the training and the inference process for image preprocessing.

\subsection{Evaluation Metrics and Baselines}
We used the area under the ROC curve (AUC), a widely-adopted metric to measure the performance of a classifier,  as our quantitative metric. ROC curve illustrates the true-positive rate versus the false-positive rate at different classification thresholds. AUC thus measures the entire two-dimensional area underneath the whole ROC curve and is a relatively fair indicator with extreme data imbalance.

In this work, we chose four advanced melanoma classification methods as baselines to ensure validity. Table.1 includes the backbone, and performance of each baseline. The implementation details of the former three baselines followed the usual practice and derived from~\cite{baseline_all}, among which we choose three models with the best performance. In addition, the lase one 'Ensemble' baseline is an ensemble of the former three baselines.

\subsection{Experimental Results}

\subsubsection{Learning Ability.}

Here we first present the results of the experiments illustrating the learning ability of our model on the final dataset (see sect.\ref{Sec_dataset}). As shown in Table\ref{tab:1}, our approach outperforms all baselines by a large extent (mean AUC: about $>$ 3.88\%). Compared with the deepest baseline among non-ensemble models (SENet-based) and the ensemble (Ensemble-based) baseline: our AUC increases by 3.40\% and 2.44\%, respectively. Besides, our model is the lightest in terms of the number of parameters (only 10.9M). The minute increase in training parameters from the IHD (0.2M) indicates that the performance improvement does not rely on additional resource consumption. This result suggests that our proposed model is superior in learning ability in terms of accuracy and computational cheapness.

\newcommand*{\Hline}[0]{%
\noalign{\global\setlength{\arrayrulewidth}{1pt}}%
\hline
\noalign{\global\setlength{\arrayrulewidth}{0.4pt}}%
}

\newcommand*{\HHline}[0]{%
\noalign{\global\setlength{\arrayrulewidth}{0.8pt}}%
\hline
\noalign{\global\setlength{\arrayrulewidth}{0.4pt}}%
}

\begin{table}
\vspace{-0.8cm}
\centering
\caption{Quantitative results on ISIC2020 dataset. 'Para': The number of learning parameters. 'BB': Our backbone. The value: mean $\pm$ standard deviation on five folds.}
\setlength\tabcolsep{4pt}
\renewcommand{\arraystretch}{1.3}
\label{tab:1}
\begin{tabular}{lcc||lcc}
\Hline
Backbone & Para(M) & AUC(\%)  & Backbone & Para(M) & AUC(\%)   \\
\HHline
Inception-v4~\cite{baseline_all} & 41.2& 86.86$\pm$1.73 & BB (EffNet-B3) & 10.7  & 92.23$\pm$0.91 \\
SENet~\cite{baseline_all} & 113.0& 89.61$\pm$1.32 & BB+GDA & 10.7 &  92.89$\pm$0.85 \\
PNASNet~\cite{baseline_all} &81.8 & 89.48$\pm$1.44 & BB+IHD & 10.9  & 92.85$\pm$0.82 \\
Ensemble~\cite{baseline_all}& 236.0 & 90.57$\pm$1.67 & \textbf{Proposed} & 10.9 & \textbf{93.01$\pm$0.59} \\
 
\Hline
\end{tabular}
\vspace{-1cm}
\end{table}

\subsubsection{Ablation Study.}This section evaluates the effectiveness of each crafted strategy to gain a deeper understanding of our approach. Specifically, we examine the AUC of both the GDA and the IHD strategies on the final dataset (see sect.\ref{Sec_dataset}). 

As shown in Table \ref{tab:1}, the mean AUC value on five-folds reaches 93.01\%. Particularly, we observe that all the 'Backbone+' settings outperform the Backbone, indicating that both the GDA and the IHD strategies make crucial contribution to excellent model performance. 

\begin{figure}
\centering
\begin{minipage}[b]{0.45\linewidth}
\centering
\includegraphics[width=\linewidth]{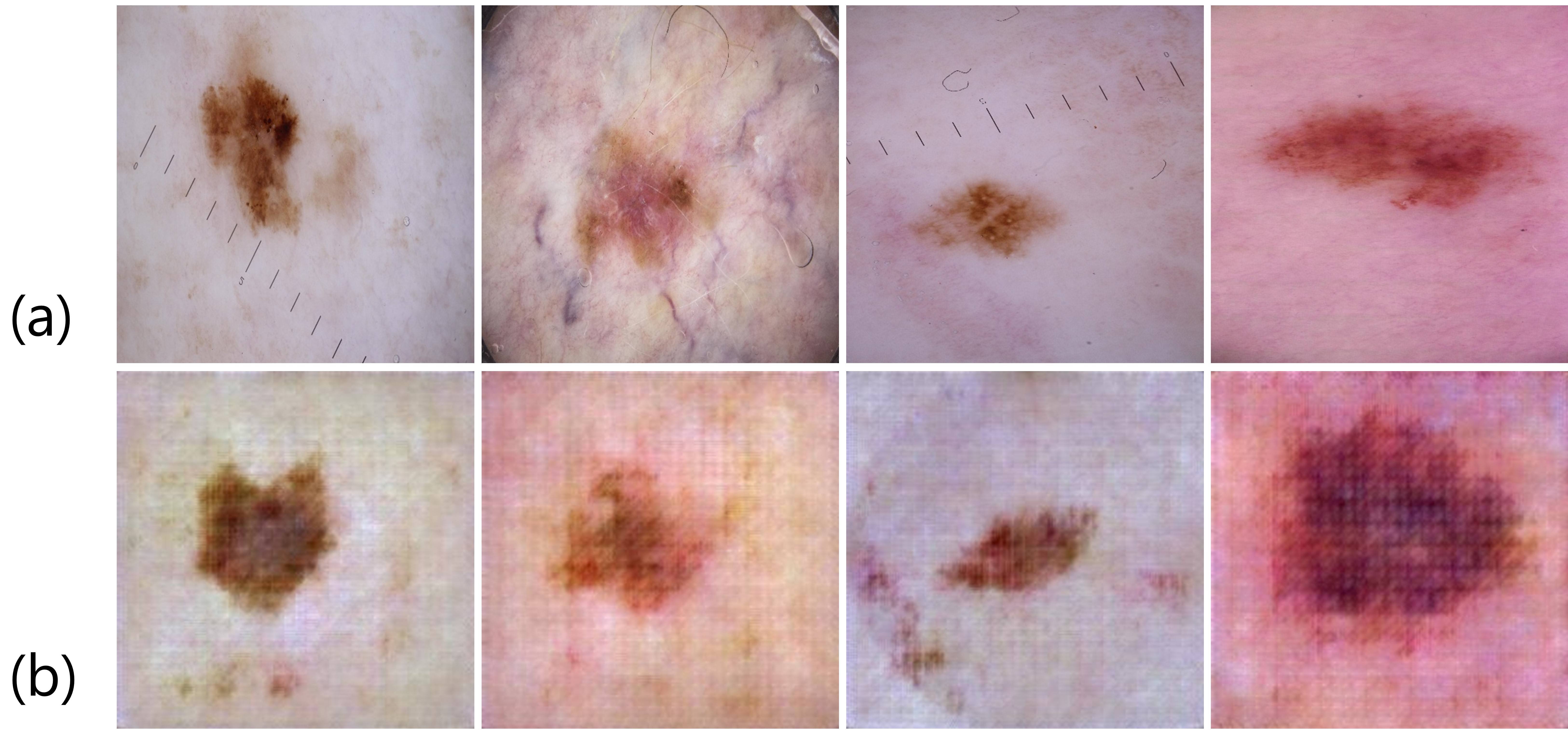}
\vspace{0.2cm}
\caption{Melanoma-positive samples. (a): original, (b): synthetic}\label{fig3}
\end{minipage}
\begin{minipage}[b]{0.54\linewidth}
\centering
\includegraphics[width=\linewidth]{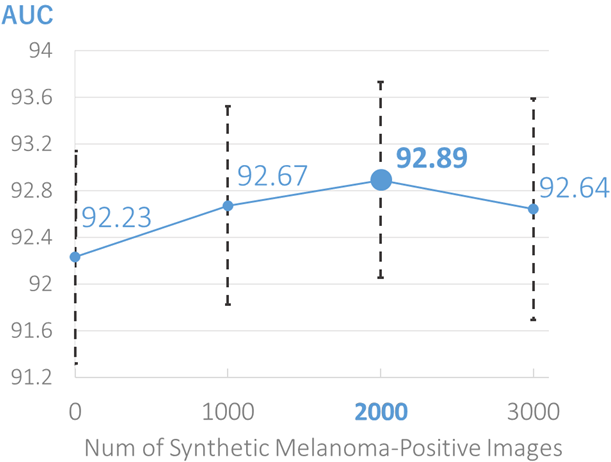}
\caption{AUC curve with standard deviation in respect of the number of synthetic images.}\label{fig4}
\end{minipage}
\vspace{-0.5cm}
\end{figure}

To further investigate the GDA component, we checked the AUC of the model using different number of synthetic melanoma-positive images (shown as Fig.\ref{fig3}) generated by the GDA. As shown in Fig.\ref{fig4}, with the increase of the number of synthetic images, the curve of AUC value rises, peaking when generating 2,000 images, and then falls, which illustrates the correlation between accuracy and man-made correction of data imbalance. Once the synthetic positive images greatly outnumber the genuine ones, the results could backfire, meaning that the model could be overwhelmed by synthetic information. The point where the AUC curve peaks (i.e., 2,000) is therefore picked as the optimized number of synthetic melanoma-positive images in our model.

To better understand our IHD strategy, we compared our IHD with preprocessing methods based on other denoising strategies. Note that, since segmentation mask-based hair removal preprocessing~\cite{GAN} requires segmented-mask as input, we adopted three morphological-based methods for comparison. Specifically, we applied each denoising approach in~\cite{hair_removal1,hair_removal2,hair_removal3} on our backbone. As shown in Table \ref{tab:2}, we found that all morphological-based preprocessing methods impose negative effects on the backbone, while our IHD brings an improvement. This result, again, firmly suggests the superiority of our IHD strategy.

\begin{figure}
\vspace{-0.5cm}
\includegraphics[width=\textwidth]{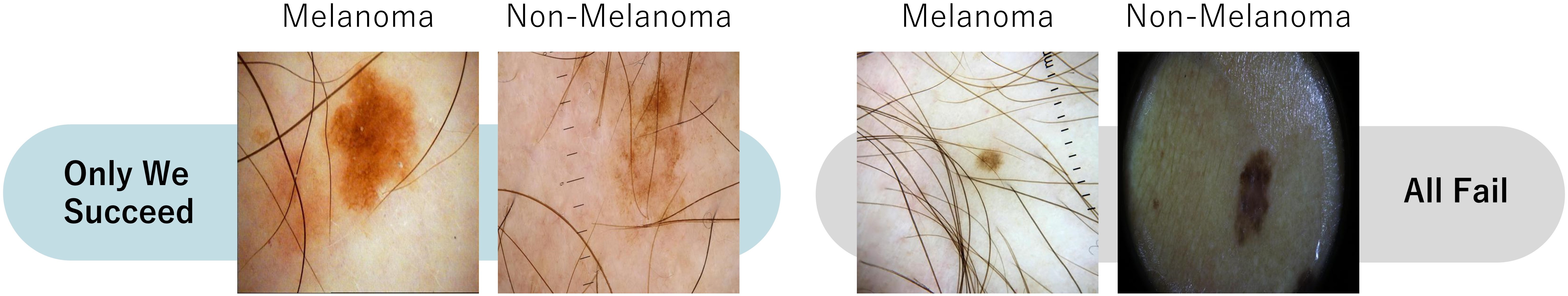}
\caption{The visualization case study. Only We Succeed: All other baselines failed to predict while ours predicts correctly. All Fail: All the models, including ours, failed to predict.} \label{fig5}
\vspace{-1cm}
\end{figure}

In addition, we provide visualization of some predictions as a case study. We picked out the cases where all baselines fail while only ours predicts correctly (annotated as 'Only We Succeed'), along with the cases where all the models, including ours, predict incorrectly (annotated as 'All Fail'). The results are shown in Fig.\ref{fig5}. Note that both images in the case of 'Only We Succeed' largely contain hair-like artifacts. This demonstrates that our approach is superior in dealing with extraneous noises and demonstrates the IHD strategy's effectiveness. As for the 'All Fail' cases, since the images are either under poor lighting or with ambiguous lesion features, this unsatisfying result could be explained by low-quality features.

\begin{table}
\vspace{-0.1cm}
\centering
\caption{Compare to other hair removal preprocessing on ISIC 2020 dataset. The value: mean value $\pm$ standard deviation on five folds.}
\label{tab:2}
\setlength\tabcolsep{20pt}
\renewcommand{\arraystretch}{1.3}
\begin{tabular}{lc}
\Hline
Methods & AUC(\%)  \\
\HHline
Our Backbone (BB)  & 92.23$\pm$0.91   \\  
BB+Hasan et al.~\cite{hair_removal1}   & 90.85$\pm$0.98   \\
BB+Bibiloni et al.~\cite{hair_removal2}  & 92.11$\pm$0.58    \\
BB+Calderon Ortiz et al.~\cite{hair_removal3} & 92.10$\pm$0.67 
\\
\textbf{BB+IHD}(Ours) & \textbf{92.85$\pm$0.82}\\
\Hline
\end{tabular}
\vspace{-0.4cm}
\end{table}

\subsubsection{Generalization Capability.}

\begin{figure}
\vspace{-0.3cm}
\centering
\begin{minipage}[b]{0.4\linewidth}
\centering
\includegraphics[width=\linewidth]{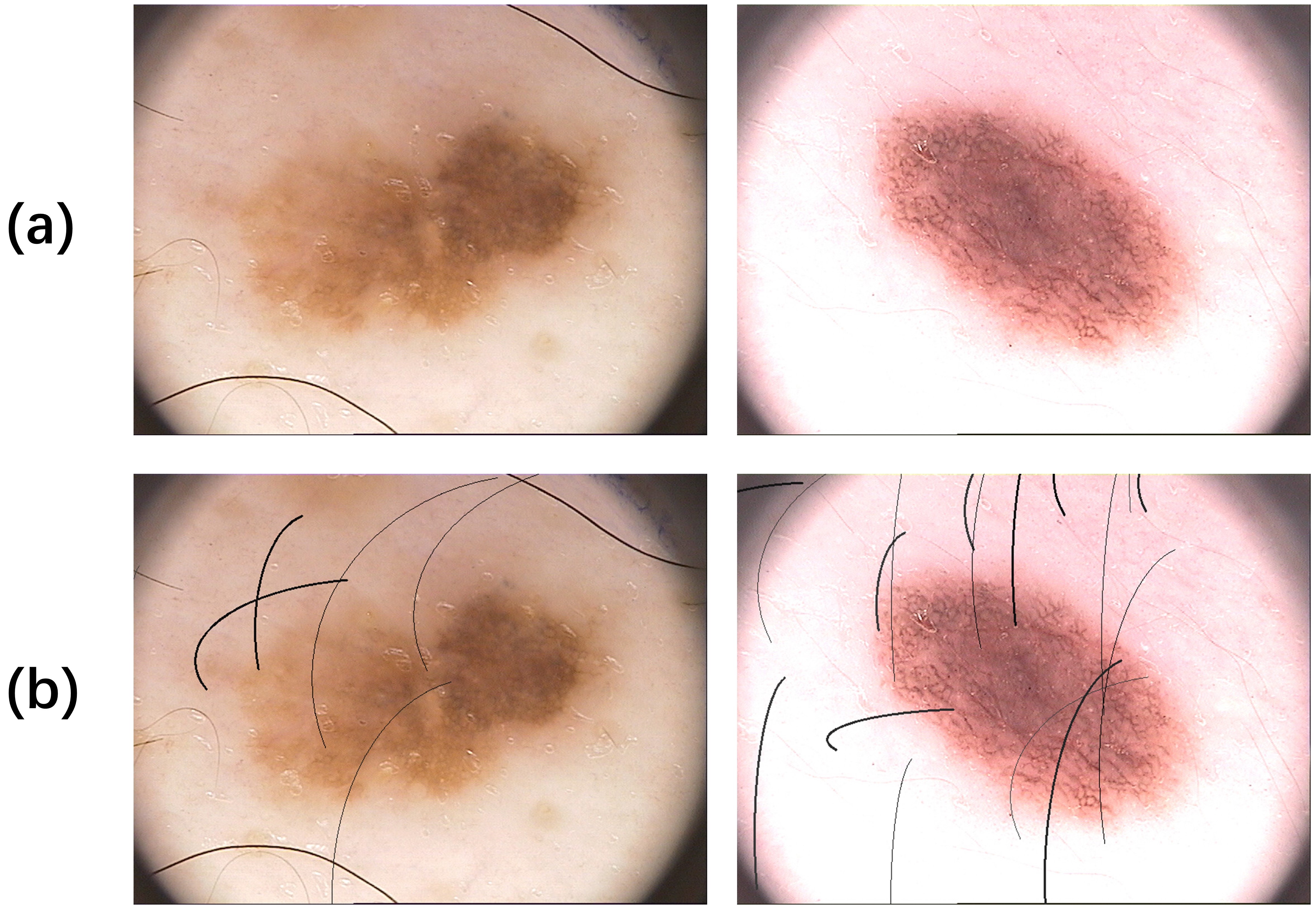}
\caption{The example of hair-augmented images from the PH$^*$. (a)original (b)augmented.}\label{fig6}
\vspace{-2cm}
\end{minipage}
\hspace{1.2cm}
\begin{minipage}[b]{0.4\linewidth}
\centering
\captionof{table}{Quantitative results on inference only dataset PH2$^*$. }\label{tab:3}
\setlength\tabcolsep{5pt}
\renewcommand{\arraystretch}{1.3}
\begin{tabular}{lc}
\Hline
Methods & AUC   \\
\HHline
Inception-v4~\cite{baseline_all} & 84.34 \\
PNASNet~\cite{baseline_all}  & 82.95   \\
SENet~\cite{baseline_all}   & 80.41     \\
Ensemble~\cite{baseline_all} & 85.66   \\
\textbf{\textbf{Proposed}} & \textbf{86.48} \\
\Hline
\end{tabular}
\end{minipage}
\vspace{-0.3cm}
\end{figure}

We evaluate the generalization of our model on the PH2 dataset~\cite{ph2}, a small dataset with only 200 dermoscopic images wherein 20.00\% are melanoma-positive and hair-like artifacts are scarce. To simulate real-world large-scale circumstances, we further augmented PH2 by adding hair noises and enlarged the PH2 to 1,000 images. The resulting dataset is annotated as PH2*. The noises are added using~\cite{addHairNoise}, which randomly added a different number of gray and black arcs to imitate hairs, as shown in Fig.\ref{fig6}.

The test results show that our model outperforms the existing advanced melanoma classifiers (see Table \ref{tab:3}), which shows that our model, empowered by the IHD strategy, has better generalization ability despite the large existence of hair-like artifacts. This result also confirms that our model shows greater potential for clinical application in large-scale datasets, wherein images are typically with high noises and of poor quality.

\section{Conclusion}
In this work, we focus on learning better lesion feature representations from dermoscopic images for melanoma classification. Wherein two main challenges: significant data imbalance, and surrounding artifacts, are observed. To resolve these challenges respectively, we propose the GAN-based data augmentation (GDA) and implicit hair denoising (IHD) strategy. Specifically, to balance the sample classes, a number of synthetic melanoma-positive images are generated through GDA and are used in the training process. Also, to implicitly eliminate hair-like artifacts, hair-related representations learned via an auxiliary network are feedbacked to and disentangled from the melanoma-feature extraction backbone. Besides, to train the IHD module, we manually labeled hair-like artifacts on top of the ISIC2020 dataset, making it the first large-scale dermoscopic dataset with annotation of hair-like artifacts. Extensive experiments indicate that our approach is capable of lesion feature representation learning, showing the promising potentials for clinical applications.The improved dataset will be publicly available after the review.

\section*{Acknowledgments}

This research was partly supported by the National Natural Science Foundation of China (Grant No. 41876098), the National Key R$\&$D Program of China (Grant No. 2020AAA0108303), and Shenzhen Science and Technology Project (Grant No. JCYJ20200109143041798). Thanks to the tutors of Special Practice of Big Data Courses, who provided valuable discussions.

%
%
%

\bibliographystyle{splncs04}
\bibliography{reference}

\end{document}